\documentclass[manuscript]{aastex}

\usepackage{graphicx}   
\usepackage{amsmath}    
\usepackage{xcolor}     

\newcommand{\pe}{p_\text{e}}

\shorttitle{ChromaStarDB}
\shortauthors{Short \& Bennett}

\begin{document}


\title{Chroma+GAS: An Expedited Solution for the Chemical Equilibrium for Cool Stellar Atmospheres}


\author{C. Ian Short}
\affil{Department of Astronomy \& Physics and Institute for Computational Astrophysics, Saint Mary's University,
    Halifax, NS, Canada, B3H 3C3}
\email{ian.short@smu.ca}

\author{Philip D. Bennett}
\affil{Department of Physics \& Atmospheric Science, Dalhousie University, Halifax, NS, B3H 4R2}
\email{philip.bennett@dal.ca}




\begin{abstract}

  We describe a unique approach to economizing the solution to the general chemical equilibrium and 
equation-of-state
problem for late-type stars, including diatomic and polyatomic molecules, that is fast, accurate, 
and suitable for responsive
approximate data modelling applications, and to more intensive modelling approaches in which the 
calculation of the gas
equilibrium must be expedited to allow other aspects to
be treated more realistically.
 The method, based on a novel economization of the Newton's method of solution of the linearized Saha
 and conservation equations,
has been implemented in
Python and made available as a stand-alone package, GASPy, and has been integrated into the interactive
Python atmosphere and spectrum modelling code ChromaStarPy.
As a result, ChromaStarPy now computes the state of the gas, the number density of absorbers, 
and the surface flux spectrum,
with consistent inclusion of 105 chemical species, including 34 diatomic, and 16 polyatomic, 
neutral molecules, as
well as H$^-$ and H$_2^+$, as well as many neutral and ionized atomic species.
The economized method converges very rapidly and greatly
improves the code's relevance to late-type stellar and brown dwarf spectrum modelling.
We provide a brief overview of
the GAS methodology, and present some illustrative results for the chemical equilibrium and spectrum
for an M-type bright giant and dwarf, and a comparison to results of the PHOENIX/PPRESS
package.
All codes are available from the OpenStars www site:
www.ap.smu.ca/OpenStars.
 
\end{abstract}



\section{Introduction}

   A proper treatment of the coupled molecular chemical equilibrium, ionization equilibrium, and equation of state (EOS) that
includes polyatomic as well as diatomic molecules is crucial for
the computational modelling of late-type stellar atmospheres and spectra for two main reasons.
1) Molecule formation, especially that of H$_{\rm 2}$, CO, and N$_{\rm 2}$ in stars
for which $N_{\rm C}/N_{\rm O} < 1$, can significantly deplete
the supply of atomic species, thus affecting the ionization equilibrium and the free electron partial pressure, $p_{\rm e}$, and
the mean molecular weight, $\mu$.
 
2) The electron pressure, needed to determine the ionization equilibrium depends, 
in cool stellar atmospheres, on a fairly large number of low-abundance, easily-ionized metal elements. 
These elements need to be included for an accurate solution of the ionization balance.
3) The strength of electronic and ro-vibrational molecular bands in the visible and near-IR emergent 
spectrum, $F_\lambda$, is sensitive to the equilibrium
concentration of trace species, some of which give rise to spectral features that are MK classification 
diagnostics, such as TiO, VO, and CH.  
Therefore, for modelling the spectral line-forming regions of G, K, and M stars
it is necessary to solve self-consistently the general coupled chemical equilibrium and EOS problem 
in the temperature range $\sim$1000 to $\sim$6000 K, and the
total gas pressure range $\sim$10 to $\sim$ 10$^5$ dyne cm$^{-2}$.

\paragraph{}

   We describe a novel, fast, accurate general procedure, GAS, for quickly computing the 
self-consistent chemical equilibrium and ionization equilibrium of a gas of typical stellar
composition, for the specified state variables of temperature and pressure. The elemental
abundances and chemical species included in the solution are specified by an 
input file listing the species, the abundance of each element, and relevant atomic 
and molecular data for each species. The current implementation of the routine in GASPy
solves the chemical equilibrium problem for 105 species,
including the first two or three ionization stages of 25 elements, H$^{-}$, 34 neutral diatomic molecules, H$_{\rm 2}^{+}$, and 16 neutral polyatomic molecules, and the corresponding
EOS.  GAS solves the completely linearized ionic and molecular Saha equations for the coupled ionization and molecular equilibrium by iterating these equations to 
convergence using the multi-dimensional Newton's method, and is a major module in the 
ATHENA stellar atmospheric modelling code.  However,
to date, GAS and ATHENA have only been described in university-archived theses 
(\citet{bennett}, \citet{bennettAthena}), although the Spectroscopy
 Made Easy (SME) package of \citet{SME1} was also based on the original GAS code of
\citet{bennett}.

\paragraph{}
There are other codes that solve the general chemical equilibrium and EOS problem, such as PPRESS (V15), the EOS module of version 15 of
the PHOENIX stellar atmosphere and spectrum modelling code
(\citet{allard95}), which is written in FORTRAN and uses the
 multi-dimensional Newton method to solve the linearized, coupled, ionic and molecular Saha equations for the partial pressures of
622 species, and
FastChem (\citet{stock2018}), written in C++, which employs a method based on decomposing the equations for the law of mass action and element
conservation into a set of coupled non-linear equations that each have one variable.  However, GAS has the advantage of being
very fast because of its unique approach to economizing the solution, and is suitable for more interactive environments such
as the Python integrated development environment (IDE), which allow a user to more quickly extract approximate results from
fitting observed spectra.  Moreover, 3D hydrodynamic atmospheric and radiative transfer codes such as that described in \citet{cobold}
must economize every other aspect of the gas equilibrium solution given the computational intensity of the problem, in
which the abundance of absorbers must be calculated at $\sim 10^6$ 3D spatial grid points.

\paragraph{}

We have ported GAS
from FORTRAN to Python and have integrated it into ChromaStarPy (CSPy, {DOI: zenodo.1095687}), an approximate general stellar atmospheric and spectrum
modelling code written in Python and described by \citet{shortbb18} and papers in that series.
CSpy, now equipped with GAS, is comparable to
the Spectroscopy Made Easy (SME) package described in \citet{SME1} and \citet{SME2}.  However, SME is a package for the
Interactive Data Language (IDL) environment that was common on Unix workstations, and pre-dates the rise of Python as a common
astronomical research environment.
The Python version of CSPy is also available as a separate
stand-alone application, GASPy.  Both codes are available from the OpenStars www site: www.ap.smu.ca/OpenStars.

\paragraph{}

In Section \ref{basics}, we provide an overview of the problem of determining the chemical equilibrium
of a gas of stellar composition that handles regimes ranging from gas temperatures cool enough for 
molecules to form, to conditions warm enough that ionization occurs. We derive the equations that
provide a general description of the state of this gas.

In Section \ref{methods} we describe the economized, linearization method for solving the chemical 
equilibrium problem, and provide an approach to obtain sufficiently accurate initial estimates to ensure
convergence of the linearization method.

In Section \ref{implementation}
we describe related improvements to CSPy that are enabled by a more realistic chemical equilibrium treatment,
in Section \ref{results} we present sample equilibrium results and a comparison to the equilibrium computed
with Phoenix and PPRESS, and in Section \ref{discussion} we describe future work suggested by this development.

\section{The GAS chemical equilibrium procedure}
\label{basics}

\subsection{Introduction and Basic Equations}

The GAS routine solves the chemical and ionization equilibrium problem for a gas of stellar composition
in thermodynamic equilibrium at temperature $T$ and pressure $p$. This requires solving the
combined molecular and ionic Saha equations for the specified chemical species and ionization
states. As an example, consider the chemical equilibrium responsible for the dissociation
of the water vapor molecule H$_2$O into its constituent atoms:
\begin{equation}
\text{H$_2$O} \rightleftharpoons 2\text{H} + \text{O}
\end{equation}
Then, the partial pressure of the constituents is related by a Saha equation of the form
\begin{equation}
K_\text{H$_2$O} = p_\text{H}^2 \, p_\text{O}/p_\text{H$_2$O}
\end{equation}
and so
\begin{equation}
p_\text{H$_2$O} = p_\text{H}^2 \, p_\text{O}/K_\text{H$_2$O}.
\end{equation}
The latter equation expresses the partial pressure of the molecular species H$_2$O in terms of
the partial pressure of the constituent neutral atoms comprising that molecule.

Similarly, consider the ionization equilibrium of carbon,
\begin{equation}
\text{C} = \text{C}^+ + \text{e}^-
\end{equation}
which implies that the partial pressure of a ``parent'' neutral species, such as C, can be 
related to the partial pressure of the singly-ionized form, C$^+$, and the 
electron pressure $p_\text{e}$, by a Saha equation of the form
\begin{equation}
I_{\text{C}^+} = p_{\text{C}^+} \, p_\text{e}/p_\text{C}
\end{equation}
and
\begin{equation}
p_{\text{C}^+} = I_{\text{C}^+} \, p_\text{C}/p_\text{e} .
\end{equation}

The equilibrium constants here, $K_\text{H$_2$O}$ and $I_{\text{C}^+}$, are functions of temperature only,
assuming the equation of state can be represented by an ideal gas. Specifically, for a neutral
atom X that ionizes to X$^+$, with the release of a free electron
\begin{equation}
\text{X} = \text{X}^+ + \text{e}^-
\end{equation}
and the ionization equilibrium constant $I_{\text{X}^+}$ has the standard Saha form given by
\begin{equation}
I_{\text{X}^+} = p_{\text{X}^+} \, p_\text{e}/p_\text{X}
     = \left( \frac{2\pi m_\text{e}kT}{h^2} \right)^{3/2} kT 
       \left( \frac{2Q_{\text{X}^+}}{Q_\text{X}} \right) \,e^{-\chi_\text{I}/kT}
\label{IX+}
\end{equation}
where $k$ is the Boltzmann constant, $h$ is the Planck constant, $m_\text{e}$ is the electron mass,
$\chi_\text{I}$ is the ionization energy of the neutral atomic species X, and $Q_\text{X}$, $Q_\text{X}^+$ 
are the internal partition functions of X and X$^+$. These partition function values are read from
a user-supplied file {\tt ``gasdata''}. This expression for the ionization equilibrium constant
can be conveniently written in logarithmic form as
\begin{equation}
\log I_{\text{X}^+} = 2.5\log T - 0.48 + \log(2Q_{\text{X}^+}/Q_\text{X}) - (5039.9/T)\chi_{\rm I}
\label{ionEqConst}
\end{equation}

These examples demonstrate the molecular and ionic Saha equations which couple the partial 
pressures of the neutral atoms and the electron pressure to the molecular and ionic partial 
pressures. We now generalize these examples to include the arbitrary molecular dissociation
and ionization equations that may occur between any constituent species of the gas in thermodynamic
equilibrium. We also develop some nomenclature necessary for this task.

In what follows the index $n$ refers to any arbitrary species in the gas other than 
free electrons, including atoms, ions, and molecules, and the index $k$ refers to the 
neutral free atomic species of element $k$.  Then the total gas pressure, $p$ is just the 
electron pressure $p_\text{e}$ and the sum of all the partial pressures $p_n$ of all the 
constituent species present in the gas:
\begin{equation}
p = p_\text{e} + \sum_n p_n
\end{equation}
Species $n$ may carry a charge, i.e., be an ion, and so we define $n'$ to be the index of the 
neutral ``parent'' species corresponding to ionic species $n$. For example,
if species $n$ is the H$_2^+$ ion, then $n'$ refers to the H$_2$ molecule.  
If $n$ is already neutral, then $n' = n$.

The equilibrium constant $I_n$ involving species $n$, as defined by the ionic Saha equation, is then
\begin{equation}
I_n = p_n \,p_{\rm e}^{q_n} /p_{n'}
\label{In}
\end{equation}
so that for any species $n$,
\begin{equation}
p_n = I_n \,p_{n'}/p_{\rm e}^{q_n}
\label{Pn_ion}
\end{equation}
where $q_n$ is the charge (ionization state) of species $n$. 

Now consider the dissociation of composite species AB into the component species A and B.
\begin{equation}
\text{AB} \rightleftharpoons \text{A} + \text{B}
\end{equation}
Here A and B may be single atoms, but may also be simpler molecules of the combined
species AB, e.g., $\text{H$_2$O} \rightleftharpoons \text{OH} + \text{H}$.

The equilibrium constant $K_\text{AB}$ for the dissociation of species A and B into the
combined species AB can be written, following the general Saha equation, as
\begin{equation}
K_\text{AB} = \frac{p_\text{A} p_\text{B}}{p_\text{AB}} =
            f(T) \frac{Q_\text{A}Q_\text{B}}{Q_\text{AB}} \,e^{-E_\text{AB}/kT}
\label{KAB}
\end{equation}
where the translational partition function, $f(T)$ is given by 
\begin{equation}
f(T) = \left( \frac{2\pi mkT}{h^2} \right)^{3/2} kT
\end{equation}
and $Q_\text{A}, Q_\text{B}, Q_\text{AB}$ are the internal functions of the respective species,
$p_\text{A}, p_\text{B}, p_\text{AB}$ are the respective partial pressures,
$m = m_\text{A}m_\text{B}/m_\text{AB}$ is the reduced mass of the combined species AB, 
and $E_\text{AB}$ is the dissociation energy into the ground states of A and B. The ionization 
equilibrium of equation~\ref{IX+} is just a special case of this with A= X$^+$ and B= e$^-$.

Now for a particle, i.e., an atom, ion, or molecule of species $n$, we define $N_{n}$ to be 
the total number of atoms present in that species. For example, if $n$ referred to water, 
H$_2$O, then $N_n= 3$, whereas if $n$ referred to neutral atomic hydrogen, H, then $N_n= 1$.  
We further define the quantity $N_{nk}$ to be the number of atoms of element $k$ present in a 
particle of species n in a molecule of species $n$. For example, again referring to species
$n$ of water, H$_2$O, and let element $k$ refer to H, then $N_{nk}= 2$, since there are 2 atoms
of H in a molecule of H$_2$O. We also let $n_k$ be the index of the $k$-th element present 
in species $n$.

Since A and B can be any species, by repeated application of equation~\ref{KAB}, the equilibrium
constant $K_n$ 
\begin{equation}
K_{n} = \left(\prod_{k}p_{{n}_{k}}^{N_{nk}}\right) / p_{n}
\label{Kn}
\end{equation}
can be found for the complete dissociation of any neutral species $n$ into its
constituent neutral atoms in terms of the molecular partition functions and dissociation energies. 
To evaluate the equilibrium constants, \citet{irwin81} fit low-order polynomials in $\ln T$ to 
the partition functions, $Q$, of molecules of astrophysical interest. The equilibrium constants, 
$K$, can then be evaluated analytically in terms of the parametrized $Q$ values. 

We adopt the simpler approach of \citet{tsuji} here and represent the values of the molecular
equilibrium constants $K_n$ by 4th-degree polynomial approximations in $\theta = 5039.9/T$. 
The necessary atomic and molecular data to determine the equilibrium constants $I_n$ and $K_n$, 
including the coefficients of the \citet{tsuji} polynomials used to approximate $K_n$,
are read from a file supplied with the GASPy code distribution.

Then, for any neutral atomic or molecular species, $n$, equation~\ref{Kn} can be solved to
obtain the partial pressure $p_n$ of species $n$
\begin{equation}
p_{n} = \left(\prod_{k}p_{{n}_{k}}^{N_{nk}}\right) / K_{n}
\label{Pn_diss}
\end{equation}
Combining this result with the ionization equation \ref{Pn_ion}, relating ionic
partial pressures to those of the neutral parent species, we obtain the partial pressure
of any species $n$ in terms of the elemental partial pressures $p_k$ and electron pressure
$p_\text{e}$.
\begin{equation}
p_{n} = {I_{n}\over K_{n}p_{\rm e}^{{\rm q}_{n}}}\prod_{k}p_{{n}_{k}}^{{\rm N}_{nk}}
\label{Pn}
\end{equation}

Finally, we define the fictitious partial pressure of $p_k^*$ to be the value of the
partial pressure $p_k$ if all molecules were fully dissociated and all atomic species
were in the neutral state. We also define the fictitious total pressure $p^*$
to be the total pressure if all molecules were fully dissociated and all atomic species
were in the neutral state. Then,
\begin{equation}
p^* = \sum_k p_k^*
\end{equation}
and for each neutral element $k$, the abundance $\alpha_k$ is
\begin{equation}
\alpha_k = p_k^*/p^* = \frac{ \sum_n N_{nk}p_n }{ \sum_n N_np_n }
\label{alphak}
\end{equation}
where $p_{k}^*$ is the fictitious partial pressure of element $k$.

We are now in a position to state the equations needed to define the numerical
problem of determining the equilibrium partial pressures of each species. Multiplying
equation~\ref{alphak} by the right-hand side denominator,
\begin{equation}
\alpha_k \sum_n N_np_n - \sum_n N_{nk}p_n = 0   \text{\hspace{0.25in} or}
\end{equation}
\begin{equation}
\sum_n (\alpha_k N_n - N_{nk})p_n = 0, ~~~~~ k=2,\cdots,K
\label{mass_bal}
\end{equation}
where the last equation \ref{mass_bal} expresses conservation of atoms of element $k$, for each 
of the total of $K$ elements considered in the equilibrium solution. The equation for $k=1$
(usually H) is omitted from the equation set because it is not linearly independent of the 
other $K-1$ equations, since
\begin{equation}
\sum_n \alpha_k = 1
\end{equation}

Charge neutrality provides another equation: the sum of all partial pressures
of charged species must be zero.
\begin{equation}
\sum_n p_nq_n - p_\text{e} = 0
\label{charge_cons}
\end{equation}

There is one final constraint: the total of all the partial pressures of the constituent
species $n$ must equal the total pressure $p$.
\begin{equation}
p_\text{e} + \sum_n p_n = p
\end{equation}
From equation \ref{charge_cons}, this becomes
\begin{equation}
\sum_n p_nq_n + \sum_n p_n = p
\end{equation}
\begin{equation}
\text{or \hspace{0.05in}} \sum_n p_n(q_n + 1) = p
\label{totalp}
\end{equation}
Equations \ref{mass_bal}, \ref{charge_cons}, and \ref{totalp} define the problem, along with
equation \ref{Pn}, which expresses the partial pressure of each species in terms of the
partial pressures of the neutral atomic elements $p_k$, for $k=2,\cdots,K$. This gives us
a total of $K+1$ equations. The unknowns are the $K$ partial pressures $p_k$ of the neutral
atomic form of the elements included in the equilibrium, and the electron pressure $p_\text{e}$,
for a total of $K+1$ unknowns. The molecular equilibrium problem is therefore well-posed.

Summary of definitions in this section:\\
$n \equiv$ index denoting arbitrary chemical species in equilibrium gas\\
$k \equiv$ index of neutral free atomic species in elemental form\\
$n_k \equiv$ index of the $k$-th element present in species $n$\\
$n' \equiv$ index of the neutral ``parent'' species of ionized species $n$\\
$p_n \equiv$ partial pressure of species $n$\\
$p_k \equiv$ partial pressure of neutral, free atomic species $k$\\
$p_k^* \equiv$ fictitious partial pressure of element $k$\\
$p_\text{e} \equiv$ electron pressure\\
$p \equiv$ total gas pressure, including electron pressure\\
$p^* \equiv$ fictitious total pressure = pressure if all species (excluding $p_\text{e}$) dissociated\\ 
$q_n \equiv$ charge (ionization state) of species n = zero for neutral species\\
$N_{nk} \equiv$ number of atoms of element $k$ in species $n$\\
$N_n \equiv$ total number of atoms in species $n$\\
$\alpha_k \equiv$ fractional abundance (by number) of element $k$ in the gas\\
$I_n \equiv$ ionization equilibrium constant of ionized species $n$\\
$K_n \equiv$ molecular equilibrium constant of molecular species $n$

\section{The Method of Solution}
\label{methods}

The GAS procedure accepts values of the state variables temperature and pressure as input, and reads other
necessary atomic and molecular data from an input file, including fractional elements abundances $\alpha_k$, 
the composition of molecular species, ionization potentials, atomic partition functions, and
coefficients of polynomial approximations of molecular equilibrium constants. The equilibrium solution
depends on powers of the elemental partial pressures $p_k$ and is inherently nonlinear. As such,
the best approach to solve the set of molecular equilibrium equations is by linearization. This requires
initial estimates be derived that are close to the exact solution, so that any differences from
this exact solution are small. Then, the linearized form of the equilibrium equations, in which small
departures from the exact solution are approximated by first-order terms, is solved. Because of the 
approximations inherent in linearization, these corrections are not exact, but if within the region of
 convergence, will yield an improved solution. In this way, this procedure can be iterated to 
convergence to the exact solution by carrying out successive solutions of the linearized equations.

Fundamentally, the equilibrium solution depends on the values of the total pressure $p$ and
electron pressure $\pe$, which for a given temperature $T$, determines the fictitious total pressure $p^*$.
Given $p^*$, the fictitious partial pressures of the elements $p_k^*$ are
given by $p_k^* = \alpha_k p^*$. Then, the partial pressure of any species $p_n$ can be found from
equation \ref{Pn}. Since $p$ and $T$ are given as input, we need to invert this solution and determine
$p^* = p^*(p,T)$ and $\pe= \pe(p,T)$. To determine reasonable initial estimates for the linearization,
we first need to obtain reasonable estimates of $p^*$ and $\pe$ at temperature $T$.

The most abundant elements dominate the fictitious total pressure $p^*$, but abundant elements $k$ 
that participate in molecule formation are inherently coupled in a nonlinear manner, so 
the main challenge is to devise a reasonably accurate initial estimate of their partial 
pressures $p_k^0$. This must be done on a case-by-case basis for the most abundant elements. 
We also need to determine an initial estimate of the electron pressure $\pe^0$, 
and this is nontrivial because at cool temperatures, the electron pressure is dominated by
contributions from several metal elements of low abundance that are easily ionized. At high
temperatures, $\pe$ is dominated by ionization of abundant elements, mostly H. We address the approach
to developing reasonable initial estimates in the next section. Note that we use a superscript ``0'' to 
indicate initial estimates of these quantities. 

\subsection{Initial Estimates of Partial Pressures}
The GAS routine obtains initial estimates of the electron and partial pressures by considering
two groups of elements: Group 1 or  ``major'' species, which contribute significantly to the 
gas pressure $p$, and Group 2 or electron donors (``metals''), which may be of low abundance but
still contribute significantly to the electron pressure $\pe$, but do not form molecules.
A few elements of low abundance that also associate into molecules are important opacity sources: these
are classified as Group 3 or ``minor'' elements, and their partial pressures can be found 
directly once $p^*$ and $\pe$ have been found from the Group 1 and 2 elements. 
An example of Group 3 species is TiO. The groups to which a particular species belongs are 
indicated in the file of atomic and molecular data read by GAS.
The group type of a species is indicated by the priority code {\tt ipr} in the input file 
of atomic and molecular data read by GAS. 

GAS assumes there are six Group 1 elements: H, C, N, O, Si and S, and specific estimates of partial 
pressures of these elements are obtained for each on a case-by-case basis.

There are nine Group 2 elements included: He, Ne, Na, Mg, Al, K, Ca, Fe and Ni.

The Group 3 elements included are, somewhat arbitrarily: Cl, Sc, Ti, V, Vr, Mn, Co, Sr, Y and Zr.

To simplify the analysis, we assume that the composition of the gas is astrophysical, i.e., the 
gas is mostly made up of H and He, with minor contributions from heavier elements. We also assume:
\begin{itemize}
\item $p_n \ll p_H^*$ for all species $n$ except for those containing H or He,
\item $p_n \ll \{p_C^*,p_N^*, p_O^*, p_{Si}^*, p_S^*\}$ for a molecular species $n$ containing a
      Group 1 element combined with {\em any} other elements, and
\item Group 3 elements do not significantly contribute to either $p^*$ or $\pe$.
\end{itemize}

Under these assumptions, the total pressure can be approximated by
\begin{equation}
p = p_{\rm H} + p_{{\rm H}_2} + p_{{\rm H}^+} + p_{\rm He} + \pe
\end{equation}
and the total fictitious pressure by
\begin{equation}
p^* \approx p_{\rm H} + 2p_{{\rm H}_2} + p_{{\rm H}^+} + p_{\rm He}
\end{equation}
so that $p^* = p + p_{{\rm H}_2} - \pe$.

To get an approximate {\em first} initial estimate of $\pe$, we consider two temperature regimes:
(1) high-temperature gas, where the source of electron pressure is dominated by the ionization of H,
and (2) low-temperature gas, where the electrons come mainly from the ionization of several metal
elements of low abundance.

For the high $T$ regime, for which $p_{\rm H_{\rm 2}}$ and $p_{\rm H^{-}} \ll p$,
and for which $p_{\rm e} \approx p_{\rm H^{+}}$, $p_{\rm e}$ can be estimated from 
the  approximation to the abundance equation for H.  Thus, for the high temperature
regime we assume
\begin{equation}
  \alpha_{\rm H} = p_{\rm H}^*/p^* = (p_{\rm H} + 2p_{{\rm H}_{\rm 2}} + p_{{\rm H}^+} +  p_{{\rm H}^-})/(p + p_{{\rm H}_{\rm 2}} - p_{\rm e}) \approx (p_{\rm H} + p_{\rm e})/(p - p_{\rm e})
\end{equation}
or
\begin{equation}
  \alpha_{\rm H}(p - p_{\rm e}) \approx p_{\rm H} + p_{\rm e} \approx p_{\rm e}(p_{\rm e}/I_{{\rm H}^+} + 1)
\end{equation}
which is a quadratic equation for $\pe= \pe^\text{hi}$, the electron pressure estimate in the 
high-temperature regime
\begin{equation}
\pe^\text{hi} \approx {1\over 2}\left[-I_{{\rm H}^+}(1+\alpha_{\rm H})
  + \sqrt{I^2_{{\rm H}^+}(1+\alpha_{\rm H})^2 + 4\alpha_{\rm H}I_{{\rm H}^+}p}\right]
\end{equation}

For the low $T$ regime, we assume
that $p_{\rm H^{-}}$, $p_{\rm H^{+}}$, and $p_{\rm e} \ll p$, and that $p_{\rm H_{\rm 2}}$ may
be significant, and that $p_{\rm e}$ is determined by the ionization state of eight low-$T$ electron donors 
that are  relatively abundant ``metals'' with modest first ground state ionization 
potentials, $\chi_{\rm I}$: C, Na, Mg, Al, Si, K, Ca, and Fe, 
so that $p^*\approx p + p_{{\rm H}_2}$ and $p_{\rm Z^+}\approx p_{\rm e}$.
For the derivation of this electron pressure estimate {\em only}, these eight electron donors are 
treated as a single fictitious element, Z, that does not form molecules, and can only ionize to
the singly-ionized state. We assume a representative ionization potential of $\chi_{\rm I,\, Z} = 7.3$ eV, 
an abundance $\alpha_{\rm Z} = \sum_k \alpha_k$, and a corresponding fictitious equilibrium constant, 
$I_{\rm Z}^+$.  The value of  $I_{\rm Z}^+$ is calculated by assuming that $Q_n=2Q_{n^+}$ so that the 
$\log(Q_{n^+}/Q_n)$ term in Eq. \ref{ionEqConst} is zero.  The electron pressure can then be estimated 
from the corresponding approximations to the abundance equations for Z and H.   
Thus, in the low temperature limit we assume for H
\begin{equation}
\alpha_{\rm H} = \frac{p_H^*}{p^*} = { (p_{\rm H} + 2p_{{\rm H}_2}) \over (p + p_{{\rm H}_2}) } = { \sqrt{K_{{\rm H}_2}p_{{\rm H}_2}} + 2p_{{\rm H}_2} \over (p_{\rm H} + p_{{\rm H}_2}) },
\end{equation}
which is a quadratic equation for $p_{{\rm H}_2}$  

\begin{equation}
p_{{\rm H}_{\rm 2}} = { { 2\alpha_{\rm H}(2-\alpha_{\rm H})p + K_{{\rm H}_{\rm 2}} } \over {2(2-\alpha_{\rm H})^2 } }
{ \bigg\{ 1 - \sqrt{ 1 - [ { {2\alpha_{\rm H}(2-\alpha_{\rm H})p} \over {2\alpha_{\rm H}(2-\alpha_{\rm H})p + K_{{\rm H}_{\rm 2}}} }  ]^2 }    \bigg\} }
\label{H2soln}
\end{equation}

For the fictitious metallic electron donor, we assume
\begin{equation}
\alpha_{\rm Z} = {p_{\rm Z}^*\over p^*} = {(p_{\rm Z} + p_{\rm Z^+}) \over p^*} \approx {p_{\rm e}\over (p + p_{{\rm H}_2})}(1 + {p_{\rm e}\over I_{\rm Z^+} })
\end{equation}
thus yielding a initial estimate for the electron pressure $\pe= \pe^\text{lo}$ in the low-temperature
regime
\begin{equation}
pe^\text{lo} \approx -{I_{\rm Z^+} + \sqrt{I^2_{\rm H} + 4\alpha_{\rm Z} I_{\rm Z}^+(p+p_{{\rm H}_2})} \over 2}
\end{equation}
where species $Z^+$ is the singly ionized stage of the fictitious metal Z, and $p_{{\rm H}_2}$ on the 
RHS is found from equation \ref{H2soln}.

Then we take $\pe = \text{max}(\pe^\text{lo}, \pe^\text{hi} )$ as our initial estimate of the electron pressure.

Under these assumptions, key fictitious partial pressures can be approximated as follows: \\
$p^*_{\rm H} = p_{\rm H} + 2p_{{\rm H}_{\rm 2}} + p_{{\rm H}^{+}} + p_{{\rm H}^{-}}$ \\
$p^*_{\rm He} = p_{\rm He}$ \\
$p^*_{\rm C} = p_{\rm C} + p_{\rm CH} + p_{\rm CO} + p_{{\rm C}^+}$ \\
$p^*_{\rm O} = p_{\rm O} + p_{\rm OH} + p_{{\rm H}_{\rm 2}{\rm O}} + p_{\rm CO} + p_{{\rm O}^+}$  \\
$p^*_{\rm N} = p_{\rm N} + p_{\rm NH} + 2p_{{\rm N}_{\rm 2}} + p_{{\rm N}^+}$ \\
$p^*_{\rm Si} = p_{\rm Si} + p_{\rm SiO} + p_{\rm SiS} + p_{\rm SiH} + p_{{\rm Si}^+}$ \\
$p^*_{\rm S} = p_{\rm S} + p_{\rm HS} + p_{{\rm H}_{\rm 2}{\rm S}} + p_{\rm SiS} + p_{{\rm S}^+}$ \\
$p^*_{\rm Cl} = p_{\rm Cl} + p_{\rm HCl} + p_{{\rm Cl}^-}$ \\
$p^*_{\rm Ti} = p_{\rm Ti} + p_{\rm TiO} + p_{{\rm Ti}^+}$   \\
$p^*_{\rm V} = p_{\rm V} + p_{\rm VO} + p_{{\rm V}^+}$   \\
$p^*_{\rm Y} = p_{\rm Y} + p_{\rm YO} + p_{{\rm YO}_{\rm 2}} + p_{{\rm Y}^+}$  \\
$p^*_{\rm Zr} = p_{\rm Zr} + p_{\rm ZrO} + p_{{\rm ZrO}_{\rm 2}} + p_{{\rm Zr}^+}$ \\

For all other elements, $k$, we calculate $p^*$ assuming only the neutral, singly-ionized, and 
perhaps, doubly-ionized stages contribute: 
$p_{k}^* = \sum_{{q_n}=0}^2 p_{k}^{(q_n)}$.

This first initial estimate of the electron pressure is refined by
iterating the linearized charge conservation equation to obtain a significantly 
improved initial estimate. Experience has shown that the converged solution 
is independent of the value chosen for $\chi_{\rm I,\, Z}$. For the linearization, 
we assume free electrons arise only from single ionizations of elements, $k$, so that

$\pe = \sum_{k} p_{{k}_+}$ and

$p_{k}^* = p_{k} + p_{{k}_+} = (1+{\pe\over{I_{k^+}}})p_{k^+}$

With $\alpha_k = p^*_k/p^*$ we have

$p_{k^+} = ( { {\alpha_kI_{k^+}} \over {I_{k^+}+\pe} }  ) p^*$

and the equation of charge neutrality can be rearranged to provide a non-linear expression 
for $\pe$ 

$\pe = \sum_k p_{k^+} = p^*\sum_k { {\alpha_kI_{k^+}}\over {I_{k^+} + \pe} }$

To clarify the dependence of $p^*$ on $\pe$ we define $\tilde{p}=p+p_{\rm H_2}$ 
so that $p^*=p+p_{\rm H_2}-\pe = \tilde{p}-\pe$ in the equation above and
\begin{equation}
\pe = (\tilde{p}-\pe) \sum_k { {\alpha_kI_{k^+}}\over {I_{k^+} + \pe} }
\label{chargeConsEst}
\end{equation}
If $\pe^0$ is a current estimate of the actual electron pressure $\pe$, 
then $\pe = \pe^0 + \delta\pe$, where we assume the correction $\delta\pe \ll \pe$.

Then linearized charge neutrality equation, equation \ref{chargeConsEst} becomes
\begin{equation}
\delta\pe = { { p^*\sum_k{ {\alpha_kI_{k^+}}\over {I_{k^+}+\,\pe^0} } - \pe^0  } 
\over 
{1 + \sum_k{ {\alpha_kI_k}\over {I_{k^+}+\,\pe^0} } + p^*\sum_k{ {\alpha_kI_k}\over {(I_{k^+}+\,\pe^0)^2} }  } }
\end{equation}
This is the linearization that we iterate to refine our initial estimate of $\pe^0$ to obtain
a value of $\pe$ consistent to first order. With this revised estimate of $\pe$, 
GAS then computes initial estimates of the partial pressures for the neutral stage 
of the elements, $p_{k}$. The number conservation equations for each Group 1 element are
based on the assumption that the molecules that these Group 1 elements participate in 
are limited to the two or three most important ones. We use initial estimates of the 
equilibrium constants defined by the ionic Saha equation ($I_{n}$)
and the molecular Saha equation ($K_{n}$).  Finally, we note that the molecular chemistry
assumed here is based on a normal stellar composition gas with $N_{\rm C}/N_{\rm O} < 1$.
Thus, for the initial estimate of $p_{\rm H}$ we assume
\begin{equation}
\alpha_{\rm H} = {p_{\rm H}^*\over p^*}\approx {p_{\rm H} + 2p_{{\rm H}_2} + p_{{\rm H}^+} +  p_{{\rm H}^-}
   \over p + p_{{\rm H}_2} - \pe }
\end{equation}
yielding
\begin{multline}
p_{\rm H} = {1 \over 2(2-\alpha_{\rm H})/K_{{\rm H}_2} }
 \left\{ -\left(1+{I_{{\rm H}^+}\over \pe} + {\pe\over I_{{\rm H}^-}}\right) +\right.\\
    \left.\sqrt{ (1+{I_{{\rm H}^+}\over \pe} + {\pe\over I_{{\rm H}^-}})^2 + 4\alpha_{\rm H}(2-\alpha_{\rm H}){(p-\pe)\over K_{{\rm H}_2}} }\right\}
\end{multline}

along with improved estimates 
$p_{{\rm H}_2} = p_{\rm H}^2 / K_{{\rm H}_2}$ and $p^* = p + p_{{\rm H}_{\rm 2}} + \pe$.

For the case of C we assume the abundance equation as follows 

\begin{equation}%
\alpha_{\rm C} = {p_{\rm C}^*\over p^*}\approx {p_{\rm C} + p_{\rm CH} + p_{\rm CO} + p_{{\rm C}^+} \over p^*}
\end{equation}
 yielding an initial estimate

\begin{equation}%
p_{\rm C} = { \alpha_{\rm C}p^* \over 1+ {p_{\rm H}\over K_{\rm CH}} + {p_{\rm O}\over K_{\rm CO}} + {I_{{\rm C}^+}\over \pe} }
\label{initialC}
\end{equation}

Then, similarly for O, we have the initial estimate

\begin{equation}%
p_{\rm O} = { \alpha_{\rm O}p^* \over 1+ {p_{\rm H}\over K_{\rm OH}} + {p^2_{\rm H}\over K_{{\rm H_{\rm 2}O}}}  +  {p_{\rm C}\over K_{\rm CO}} + {I_{{\rm O}^+}\over \pe} }
\label{initialO}
\end{equation}

We then substitute Eq. \ref{initialC} into Eq. \ref{initialO} to solve for the initial estimate of $p_{\rm O}$, and then back-substitute into Eq.
\ref{initialC} to solve for the initial estimate of $p_{\rm C}$.

A similar set of two abundance equations for two unknowns is set up for Si and S, with Eq. \ref{initialO} substituted for $p_{\rm O}$ in the equation for Si.
For N the abundance equations leads to

\begin{equation}%
{2p^2_{\rm N}\over K_{{\rm N_{\rm 2}}}} + p_{\rm N}(1+ {p_{\rm H}\over K_{\rm NH}} + {I_{{\rm N}^+}\over \pe}) - \alpha_{\rm N}p^* = 0
\end{equation}

 which is a quadratic equation for the initial estimate of $p_{\rm N}$.  Similar abundance equations can be solved for the
initial estimates of the Group 3 elements that bond to H or O, including Ti, V, Y, and Zr, substituting the value of 
$p_{\rm O}$ from Eq. \ref{initialO} into their equations, and for Cl, which depends on the value of $p_{\rm H}$. 

For any Group 2 elements for which an initial estimate is needed, for this purpose we assume the element is present in only the neutral or singly ionized form so that 

\begin{equation}
p_{n} = {\alpha_{n}p^*\over 1 + I_{{n}^+}/\pe }
\end{equation}

With this, we have obtained initial estimate of the gas and electron partial pressures
that should be sufficiently accurate for the main linearization solution to converge.

\subsection{The Linearized Solution of the Economized Equations}

The execution time required for the solution scales as $n^2$, where $n$ is the number of
species included in the chemical equilibrium.  Therefore, GAS implements an
''economized'' solution based on a fictitious ``metallic'' element Z that represents those
elements that only contribute electrons via ionization and are not significantly 
involved in molecular association: these are the Group 2 elements.
GAS solves eight coupled linearized equations for the eight first order corrections:
$\delta p_\text{H}$, $\delta p_\text{C}$, $\delta p_\text{N}$, $\delta p_\text{O}$, 
$\delta p_\text{Si}$, $\delta p_\text{S}$, $\delta p_\text{Z}^*$, $\delta\pe$.

The eight equations to be solved are the linearized versions of the five abundance 
equations for C, N, O, Si, S, and the fictitious metal Z, each of which expresses the
constraint that the abundance of each element $k$ is consistent with $\alpha_k=p_k^*/p^*$.
One of the abundance equations
is linearly dependent on the rest and should not be included. We have chosen to eliminate 
the abundance equation for H from the set. The remaining two equations are
the total pressure equation: the sum of all the partial pressures must equal the total
pressure, and the charge neutrality equation: the sum of all charge must be zero.

The abundance equation (equation \ref{massBalance2})  for each major (Group 1) element $k$ is

\begin{equation}
\sum_{n} (\alpha_{k}N_{n} - N_{nk})p_{n} + \alpha_{k}p^*_{\rm Z} = 0
\label{massBalance2}
\end{equation} 

The initial estimate, $p_n^0$, obtained by the methods of the previous section should be close
to the exact solution $p_n$. We can write the exact solution $p_n = p_n^0 + \delta  p_n$,
where the correction term $\delta p_n$ is assumed small $(\delta p_n/p_n \ll 1)$. Then, the equation can be
written in terms of the small corrections, and only terms of first order in the correction terms
kept. In this way, the equation is linearized, and the linear system solved for the corrections
$\delta p_n$, which are then used to update the solution. 

After $i$ iterations, we obtain an estimate of the partial pressure $\delta p_n^{i+1}$, which can
be added to the previous solution to obtain an improved estimate
\begin{equation}
p_n^{i+1} = p_n^i + \delta p_n^{i+1}
\end{equation}
This process is iterated repeatedly until the changes become small, and the system converges 
to the exact solution $p_n$.
 
The linearized version of equation \ref{massBalance2} is
\begin{equation}
\sum_{n} (\alpha_{k}N_{n} - N_{nk})\delta p_{n} + \alpha_{k}\delta p^*_{\rm Z}
 = \sum_{n} (N_{nk} - \alpha_{k}N_{n})p_{n} + \alpha_{k}p^*_{\rm Z}
\label{linearAbund}
\end{equation}
where the superscript iteration number ``$i$'' has been dropped on the partial pressures estimates 
for clarity.

This provides equations for the corrections $\delta p_{n}$ for {\it all species}, including molecules and ions, 
whereas our independent variables are the eight neutral Group 1 elements and $\pe$.  To relate $\delta p_{n}$
to $\delta p_{n_k}$ and $\delta_{\rm e}$ we linearize Eq. \ref{Pn} as follows

\begin{equation}
p_{n} + \delta p_{n} = {I_{n}\over K_{n}(\pe+\delta \pe)^{{\rm q}_{n}}}\prod_{k}(p_{n_k}+\delta p_{n_k})^{{\rm N}_{nk}}
\end{equation}

which can be rearranged such that

\begin{equation}
\delta p_{n} = p_{n}(\sum_k { {N_{nk}} \over {p_{n_k}} }\delta p_{n_k} - q_n{ {\delta \pe} \over {\pe} }  )
\label{deltaPn}
\end{equation}

assuming that $\delta p_{n_k}/p_{n_k} \ll 1$ and $\delta \pe/\pe \ll 1$.  Substituting the above
into Eq. \ref{linearAbund} gives us the linearized mass balance equations for the eight Group 1 elements 

\begin{multline}
\sum_{n} (\alpha_{k}N_{k} - N_{nk})p_{n}
 \sum_{k} {N_{nk}\over p_{{n}_{k}}}\delta p_{{n}_{k}}\\
  - {1\over \pe}\left[\sum_{n} (\alpha_{k}N_{n} - N_{nk})p_{n}q_{n}\right]\delta \pe
  + \alpha_{k}\delta p_{\rm Z}^* \\
  = \sum_{n} (N_{nk} - \alpha_{k}N_{n})p_{n} - \alpha_{k}p_{\rm Z}^*
\label{abndnc}
\end{multline}

where $k$ ranges over Group 1 elements only, and $n$ refers only to species formed entirely from Group 1 elements. 

Similarly, the linearized abundance equation for the fictitious metal $Z$ is

\begin{multline}
\alpha_{\rm Z}\sum_{n} N_{n}p_{n}
  \sum_{k}{N_{nk}\over p_{{n}_{k}}}\delta p_{{n}_{k}}\\
  - {\alpha_{\rm Z}\over \pe}(\sum_{n}N_{n}p_{n}q_{n})\delta \pe
  - (1-\alpha_{\rm Z})\delta p_{\rm Z}^*\\
  = -\alpha_{\rm Z}\sum_{n}N_{n}p_{n} + (1-\alpha_{\rm Z})p_{\rm Z}^*
\label{zabndnc}
\end{multline}


The total pressure equation (equation \ref{totalPressure2}) is now
\begin{equation}
\sum_{n} p_{n} + p_{\rm Z} + p_{\rm Z^+} + \pe  = p
\label{totalPressure2}
\end{equation}
  
With $p_n = p_n^0 + \delta p_n$, $\pe = \pe^0 + \delta\pe$, 
and $p^*_{\rm Z} = p^{*0}_{\rm Z} + \delta  p^*_{\rm Z}$ this linearizes to 

\begin{equation}
\sum_{n} \delta p_{n} + \delta p_{\rm Z} + \delta p_{\rm Z^+} + \delta p{\rm e}  = p - \sum_{n} p_{n} 
         - p_{\rm Z} - p_{\rm Z^+} - \pe
\label{totalPressure2}
\end{equation}
where the superscript ``0''s have been dropped on the initial partial pressure estimates for clarity.
Substituting equation \ref{deltaPn} for $\delta p_n$ again, the linearized total pressure equation is

\begin{multline}
\sum_{n}p_{n}\sum_{k}{N_{nk}\over p_{{n}_{k}}}\delta p_{{n}_{k}}
  + (1-{1\over \pe}\sum_{n}p_{n}q_{n})\delta \pe
  + \delta p_{\rm Z}^*\\
   = p - \sum_{n}p_{n} - p_{\rm Z}^* - \pe
\label{totpres}
\end{multline}

The modified charge neutrality equation is  

\begin{equation}
\sum_{n} p_{n}q_{n} + p_{\rm Z^+} - p_{\rm e} = 0
\label{chargeNeutral2}
\end{equation}

With $p_n = p_n^0 + \delta p_n$, $\pe = \pe^0 + \delta\pe$, 
and $p^*_{\rm Z} = p^{*0}_{\rm Z} + \delta  p^*_{\rm Z}$ this linearizes to 

\begin{equation}
\sum_{n} q_n\delta p_{n} + \delta p_{\rm Z^+} - \delta \pe = \pe - \sum_{n} q_n p_{n} - \delta p_{\rm Z^+}
\label{chargeNeutral2}
\end{equation}
where the superscript ``0''s have been dropped on the initial partial pressures estimates for clarity.
This result  must be expressed in terms of the independent variable $p^*_{\rm Z}$, which we do as follows

\begin{equation}
p_{\rm Z^+} = \sum_m p_{m^+} = p^*\sum_m { {\alpha_m I_{m^+}} \over {I_{m^+} + \pe} } 
=  {{p^*_{\rm Z}} \over {\alpha_{\rm Z}} } \sum_m { {\alpha_m I_{m^+}} \over {I_{m^+} + \pe} } 
\end{equation}

so that the modified charge neutrality equation is now

\begin{equation}
\sum_{n} p_{n}q_{n} + {{p^*_{\rm Z}} \over {\alpha_{\rm Z}} } \sum_m { {\alpha_m I_{m^+}} \over {I_{m^+} + \pe} } = \pe
\end{equation}

again, where the superscript ``0''s have been dropped. 
Substituting Eq. \ref{deltaPn} for $\delta P_n$ again, 
the charge neutrality equation (Eq. \ref{chargeNeutral2}) can be linearized to

\begin{multline}
\sum_n p_nq_n\sum_k{N_{nk}\over p_{{n}_{k}}}\delta p_{n_k} - 1 \\
+ {1\over \pe}\sum_np_nq^2_n
 + {p_{\rm Z}^*\over\alpha_{\rm Z}}\sum_m{\alpha_m I_m^+\over (I_{m^+}+\pe)^2 }\delta \pe\\
  + {1\over\alpha_{\rm Z}}\left(\sum_m {\alpha_m I_m^+}\over {(I_m^+ +\pe)} \right)\delta p_{\rm Z}^* \\
  = -\sum_n p_n q_n
  - {p_{\rm Z}^*\over \alpha_{\rm Z}}\sum_m {\alpha_m I_m^+\over (I_m^+}+\pe)
  + \pe
\label{charge}
  \end{multline}

where species $m^+$ are the singly ionized stages of the metals $m$ contributing to the fictitious
metal $Z$.

 
We have a set of eight equations for eight unknowns:  Equation \ref{abndnc} for five of the Group 1 elements
(we are over-constrained by
one equation and omit the equation for $\delta p_{\rm H}$), Eq. \ref{zabndnc}, Eq. \ref{totpres}, and
Eq. \ref{charge}, which we solve for the eight unknown corrections: 
$\delta p_{\rm C}$, $\delta p_{\rm N}$, $\delta p_{\rm O}$, $\delta p_{\rm Si}$, $\delta p_{\rm S}$, 
$\delta p^*_{\rm Z}$, and $\delta \pe$.
The system is solved using the LINPACK procedure
DGEFA to factorize the full matrix of coefficients, $a$, and to reduce it to upper triangular form by Gaussian elimination, 
and then
the LINPACK procedure DGESL is used with $a$ to solve for the corrections $\delta X$.  If any of the diagonal 
elements of the upper triangular factor of $a$ are zero, DGEFA will return the corresponding array subscript 
along with the other outputs, allowing us to detect cases where DGESL will divide by zero.
Python implementations typically 
represent floating-point numbers
as double-precision by default (64-bit).  Currently, if the procedure does not achieve the convergence criterion
within ten iterations, it will print a warning to the standard output. 

Once the values of $p_{\rm Z}^*$ and $\pe$ are converged,
we recover the $p$ values for the individual metals from their input abundances, $\alpha$, as follows

\begin{equation}
 p_m = ({\alpha_m\over \alpha_{\rm Z}}){\pe p_{\rm Z}^*\over (I_{m^+}+\pe)}
\end{equation}

\subsection {Input }

The GAS procedure \citep{bennett} takes as input the equilibrium gas temperature $T$ and the 
total gas pressure, $p$.
Additionally, the code reads an arbitrary list of atomic, ionic, and molecular species, $n$, from a 
user-supplied file (``{\tt gasdata}''), and the list should include the six elements that most strongly 
couple to the molecular equilibrium: H, C, N, O, Si, and S.  The species-wise records specify
the chemical
symbol of the species (name$[\, ]$), the ''priority code'' equal to 1, 2, or 3, indicating how that species is to be included in the
treatment (ipr$[\, ]$), the electronic charge in charge units (nch$[\, ]$), the total number of different elements
comprising that species (nel$[\, ]$), and one or more pairs of values specifying the number of atoms of each element (nat$[\, ]$) and
the corresponding atomic number of that element (zat$[\, ]$) that comprises that species.  If the species is a neutral atom then the
record includes the total abundance of that element (in all its forms), $\alpha_{k}$ as defined above, (comp$[\, ]$),
thus specifying the input chemical composition,
and the atomic weight in amu (awt$[\, ]$).
 If the species is an ion then the record includes the ground state ionization energy from the
 next lowest ionization stage in eV (ip$[\, ]$), followed by the term $\log 2Q_{X^+}/Q_X$ appearing
in the expression for the logarithm of the ionization constant ($\log I_{X^+}$) in equation \ref{IX+}.
The values of that ratio of partition functions in the {\tt gasdata} file are for $T=5040$ K, from 
\citet{allen73}.   
If the species is a molecule, then the record includes the five coefficients of the quartic polynomial fit to the
equilibrium constant,  $K_n(T)$, defined by the molecular Saha equation for species $n$, as a function of temperature
(\citet{tsuji}) (logk$[\, ]$).  There are currently 105 records covering all of the species included, and this arrangement
allows new species to be added to the treatment {\it ad hoc} by adding records to the input file.

\subsection{Output and performance}

GAS produces consistently calculated values for $\rho$, $\pe$, $\mu$, and the $p_{n}$ values for all 105
species, $n$, currently being included.
The convergence criterion is $(p^{\rm i}_{n} - p^{{\rm i}-1}_{n})/p^{\rm i}_{n} < \epsilon$ for all six
Group 1  species, the fictitious metal $Z$, and $e^-$ particles.  In ATHENA $\epsilon$ is set to $10^{-4}$ and
the GAS procedure typically converges in 2 to 3 iterations for stars of $N_{\rm C}/N_{\rm O} < 1$, and for stars of any
$N_{\rm C}/N_{\rm O}$ value if $T_{\rm kin}(\tau) > 3000$ K at all $\tau$.  Many more iterations may be required
for stars of $N_{\rm C}/N_{\rm O} > 1$ and where $T_{\rm kin}(\tau) < 3000$ K for some $\tau$ range because the
starting approximation currently assumes that Group 1 elements
are only depleted by molecules that dominate in an O-rich composition.  Because
CSPy is intended for rapid responsiveness that is just realistic enough for initial and demonstrative data modeling,
we set $\epsilon$ to $10^{-2}$.  In practice we find that replacing the previous ionization equilibrium and EOS
procedure in CSPy with GAS has a negligible effect on wall-clock time.

\section{Implementation in CSPy}
\label{implementation}

   Previously, CPy arrived at values of $\pe(\tau_{\rm Ros})$, $\rho(\tau_{\rm Ros})$,
$\mu(\tau_{\rm Ros})$, and $p_{n}(\tau_{\rm Ros})$ for atomic species by a straightforward iteration of
the coupled ionic Saha equations starting from an initial guess at $\pe(\tau_{\rm Ros})$ computed with
the method described in \citet{gray}.  For stars of $T_{\rm eff} < 5000$ K the value of $p_{\rm TiO}(\tau_{\rm Ros})$
was then computed {\it post facto} and all other molecules were neglected.  This is still the procedure for stars of
$T_{\rm eff} > 6500$ K.

\paragraph{}

   For stars of $T_{\rm eff} < 6500$ K, CSPy now calls the GAS procedure to obtain the values of $\pe(\tau_{\rm Ros})$,
$\rho(\tau_{\rm Ros})$,
$\mu(\tau_{\rm Ros})$, and $p_{n}(\tau_{\rm Ros})$ for all neutral and singly ionized atomic species (and doubly ionized species
for Mg and Ca), H$^-$, and all molecular species that are accounted for in GAS at all Rosseland optical depths $\tau_{\rm Ros}$.  The call to GAS
is part of an iterative procedure that includes calculation of the monochromatic ($\kappa_\lambda(\tau)$) and Rosseland
mean ($\kappa_{\rm Ros}(\tau)$) mass extinction coefficients,
and integration of the hydrostatic equilibrium equation (HSE) on the $\tau_{\rm Ros}$ scale to
improve the estimates of $p(\tau_{\rm Ros})$ and $p(\tau_{\rm Ros})$.  CSPy codes then evaluate the ionic Saha
equation to obtain values for $p_{n}(\tau_{\rm Ros})$ for any atomic species not accounted for in GAS.

\subsection{Improvements to molecular opacity}

\subsubsection{TiO opacity}

  Now that we can compute much more realistic values of $p_{\rm TiO}$, we have increased the number of TiO bands that we include in the
Just Overlapping Line Approximation (JOLA) \citep{jola} in the
computation of the emergent synthetic spectrum.  In addition to the original $C^3\Delta -X^3\Delta$ ($\alpha$ system, $\omega_{\rm 00} = 19341.7$ cm$^{-1}$),
$c^1\Phi -a^1\Delta$ ($\beta$ system, $\omega_{\rm 00} = 17840.6$ cm$^{-1}$) , and
$A^3\Phi -X^3\Delta$ ($\gamma$ system, $\omega_{\rm 00} = 14095.9$ cm$^{-1}$)
systems that we were already including \citep{shortbb18}, we now also include the $B^3\Pi -X^3\Delta$ ($\omega_{\rm 00} = 16148.5$ cm$^{-1}$),
$E^3\Pi -X^3\Delta$ ($\omega_{\rm 00} = 11894.0$ cm$^{-1}$),
$b^1\Pi -a^1\Delta$ ($\omega_{\rm 00} = 11272.8$ cm$^{-1}$), and $b^1\Pi -d^1\Sigma$ ($\omega_{\rm 00} = 9054.0$ cm$^{-1}$) systems.
The molecular data for the four newly added systems is from \citet{jorgensen}.
Frustratingly, we continue to have to tune, {\it ad hoc}, the unknown ``line strength'' factor, $S$, in the calculation of the
band oscillator strength (see \citet{allens}), and an honest description of the procedure should acknowledge that.
The addition of these four bands allows the overall spectral energy distribution (SED) of M stars to be more realistic. 
Now that we have incorporated the GAS package, the way is open to adding many more JOLA bands to
represent other important molecular absorption features, including the CH $\lambda 4300$ G band, which is another
important molecular MK classification diagnostic.

\subsubsection{Rayleigh scattering}

 GAS allows us to compute for the first time in CSPy the value of $p_{\rm H_{\rm 2}}$ and allows us to now compute the contribution of
$H_{\rm 2}$ Rayleigh scattering to the total continuous extinction coefficient, $\kappa^{\rm C}_\lambda(\tau)$.  As described in \citet{csdb17}, we compute
the contribution to Rayleigh scattering opacity for all sources with the routines ported from the Moog spectrum synthesis code \citep{moog}.

\section{Results}
\label{results}

\begin{figure}
\includegraphics[width=\columnwidth]{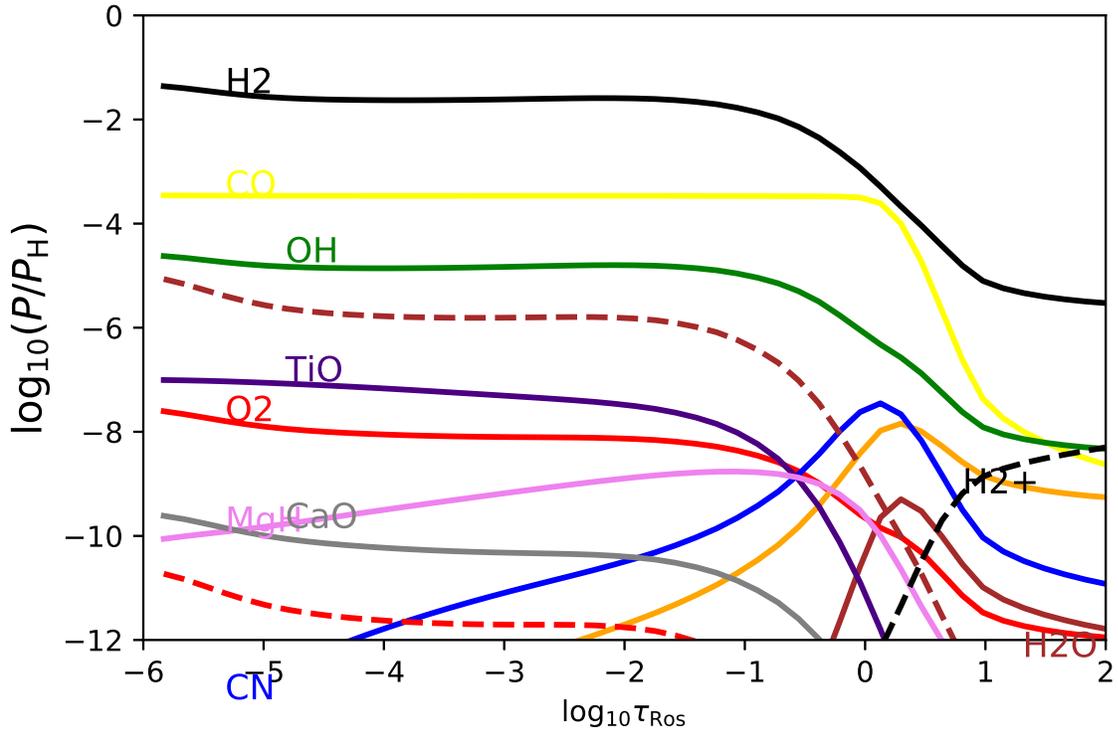}
\caption{Partial pressures for select molecular species with respect to $p_{\rm H}$ as a
function of Rosseland mean optical depth for a solar metallicity model of $T_{\rm eff} = 3600$ K
and $\log g = 1.0$, representative of a bright M giant ($N_{\rm C}/N_{\rm O} < 1$) with strong TiO bands.
  \label{fPPlogg1}
}
\end{figure}

\begin{figure}
\includegraphics[width=\columnwidth]{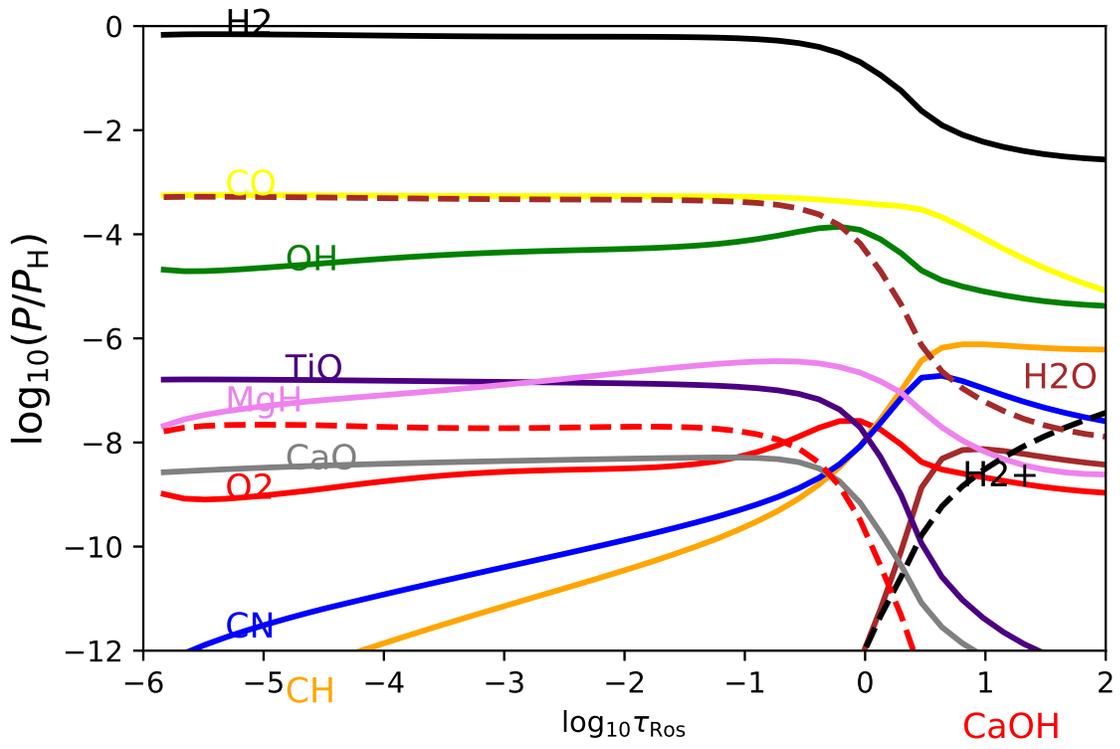}
\caption{Same as Fig. \ref{fPPlogg1}, but for a model of $\log g = 5.0$, representative of a
very late-type M dwarf.
  \label{fPPlogg5}
}
\end{figure}

\begin{figure}
\includegraphics[width=\columnwidth]{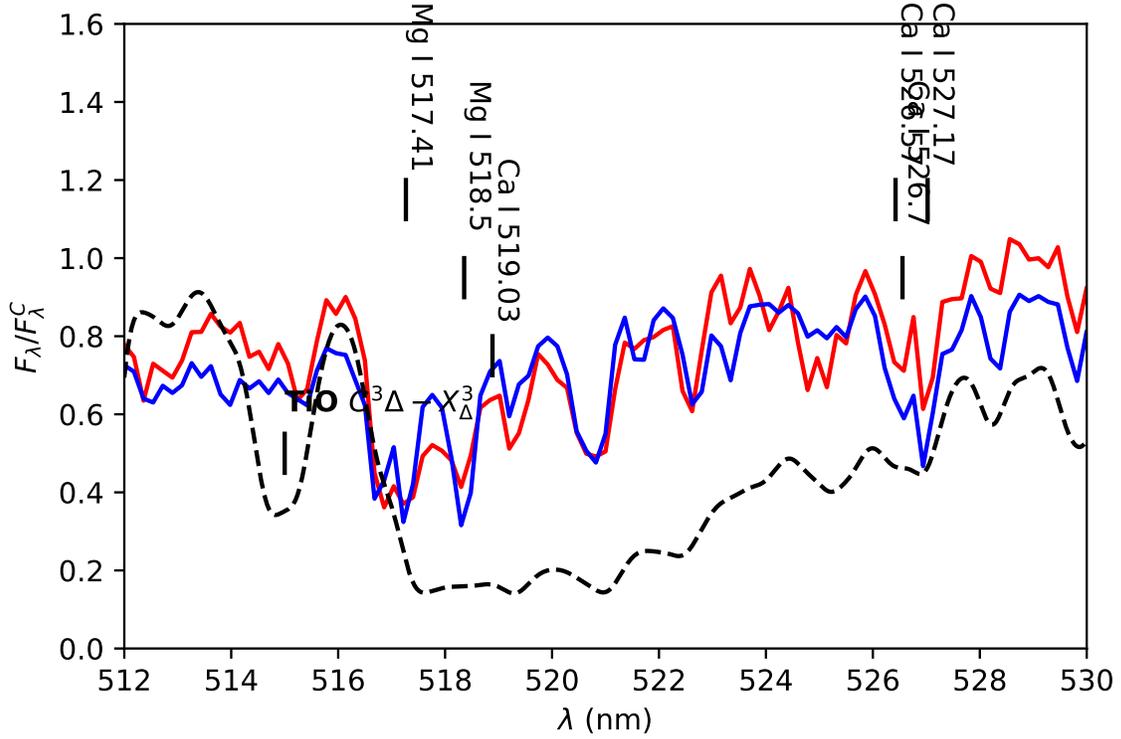}
\caption{Region around the strongest absorption of the TiO $\alpha$ system electronic
band, $C^3\Delta -X^3\Delta$ ($\omega_{\rm 00} = 19341.7$ cm$^{-1}$).  Black line:
Synthetic spectrum for a solar metallicity model of $T_{\rm eff} = 3600$ K
and $\log g = 1.0$, representative of a bright M giant ($N_{\rm C}/N_{\rm O} < 1$), broadened to the
nominal instrumental resolution of the MILES spectrograph ($\Delta\lambda = 0.25$ nm).
Red line:  Observed spectrum from the MILES library for a star with
stated parameters in the MILES catalogue of $T_{\rm eff} = 3600$ K, $\log g = 1.10$,
 and $[{\rm Fe}/{\rm H}] = +0.02$ (HD007351, MILES catalogue spectral class M2 and designation s0053).
Blue line:  As for the red line,
but for a star of stated $T_{\rm eff} = 3600$ K, $\log g = 0.80$,
 and $[{\rm Fe}/{\rm H}] = -0.19$ (HD147923, MILES spectral class M and designation s0593).
 A single-point
renormalization factor of 0.8 was applied, {\it ad hoc}, to the MILES spectra.
  \label{fTiOlogg1}
}
\end{figure}

\begin{figure}
\includegraphics[width=\columnwidth]{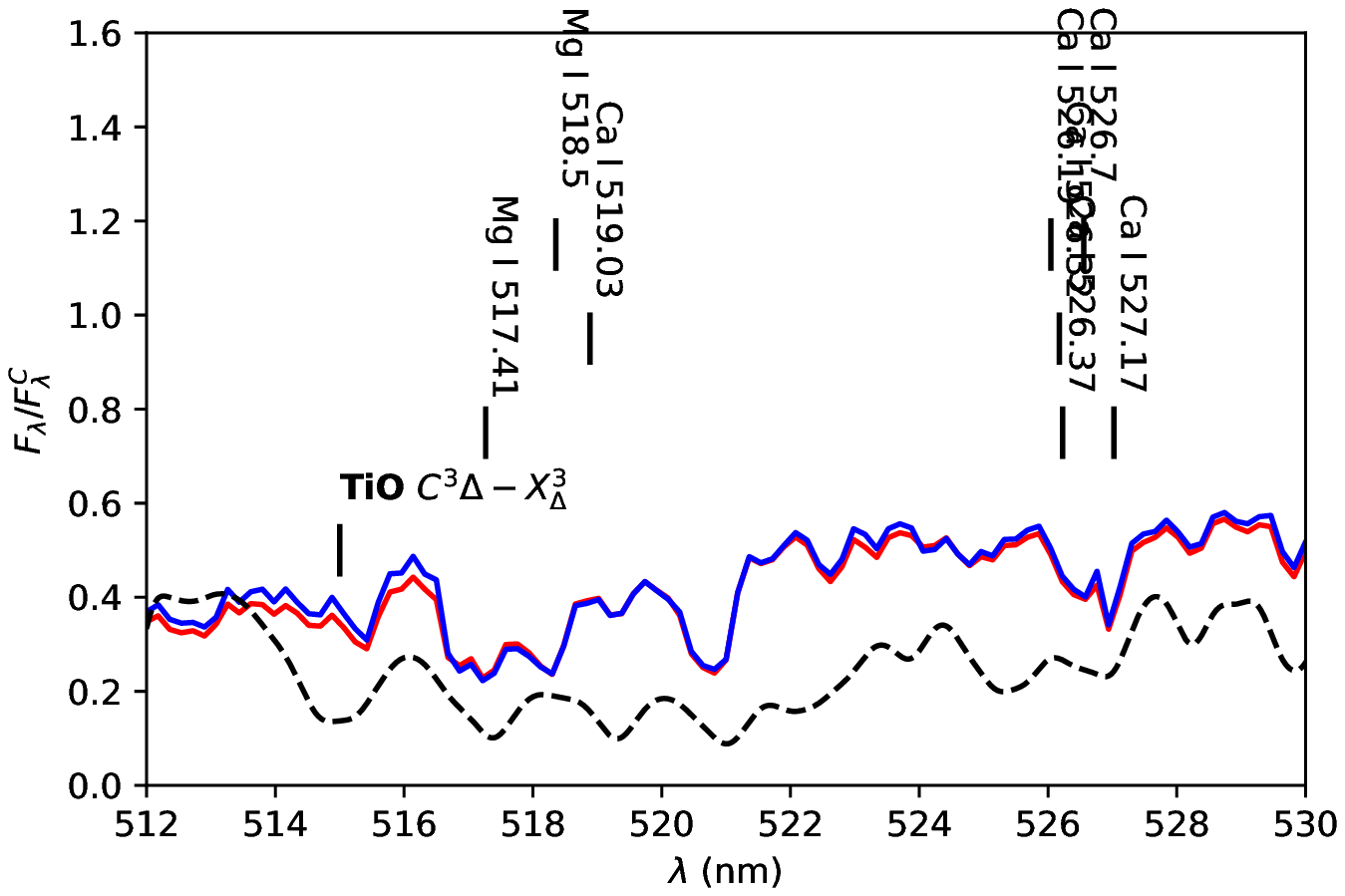}
\caption{Same as Fig. \ref{fTiOlogg1}, but for a model of $\log g = 5.0$, representative of a
very late-type M dwarf, and observed MILES stars of stated $T_{\rm eff} = 3620$ K, $\log g = 4.93$,
 and $[{\rm Fe}/{\rm H}] = -1.50$ (BD+442051A, MILES catalogue spectral type M2 V and designation s0399, red line), and
$T_{\rm eff} = 3551$ K, $\log g = 4.90$,
 and $[{\rm Fe}/{\rm H}] = -0.20$ (HD095735, MILES catalogue spectral type M2 V and designation s0398, blue line).
 A single-point
renormalization factor of 0.5 was applied, {\it ad hoc}, to the MILES spectra.
  \label{fTiOlogg5}
}
\end{figure}

In Figs. \ref{fPPlogg1} and \ref{fPPlogg5} we present partial pressure values with respect to that of H for a selection of the most important molecules,
including TiO, throughout the atmosphere
for, respectively, an evolved (low $\log g$) and an un-evolved (high $\log g$) M star of $T_{\rm eff} = 3600$ K and solar
metallicity ($N_{\rm C}/N_{\rm O} < 1$).  In Figs. \ref{fTiOlogg1}
and \ref{fTiOlogg5} we show the surface flux spectrum, $F_\lambda$, in the region of the strongest absorption caused by the
TiO $\alpha$ system electronic
band, $C^3\Delta -X^3\Delta$ ($\omega_{\rm 00} = 19341.7$ cm$^{-1}$) for the same two models, and the comparison to relevant
observed spectra taken from the MILES library (\citet{miles1}, \citet{miles2}).  Abundances are those of
\citet{grevs98}.

\paragraph{}

We were only able to find two M dwarfs ($\log g > 4.5$) in the MILES library
with a catalogue value of $T_{\rm eff} = 3600 \pm 500$ K, and one has $[{\rm Fe}/{\rm H}] = -1.50$ (BD+442051A, MILES catalogue
number s0399).
However, the TiO band in
our synthetic spectrum is computed with the JOLA approximation, and an approximate {\it ad hoc} tuning of the line strength
parameters, $S$, so we view the comparison as useful despite the discrepancy in $[{\rm Fe}/{\rm H}]$ values.   The other M dwarf
(HD095735, MILES catalogue number s0398) is closer to solar metallicity with $[{\rm Fe}/{\rm H}] = -0.20$.  We convolved our
synthetic spectrum with a Gaussian kernel of FWHM equal to the nominal spectral resolution of the MILES spectrograph,
corresponding to $\Delta\lambda = 0.25$ nm.  Given the moderate resolution and large $\lambda$ range of our comparison,
and the approximate nature of the JOLA treatment of molecular band opacity, we did not apply any other corrections to the
synthetic spectrum, and its $\lambda$ scale is that of rest wavelength in vacuum in the star's centre-of-mass frame.

\subsection{Comparison to PHOENIX V15 and PPRESS}

\begin{figure}
\includegraphics[width=\columnwidth]{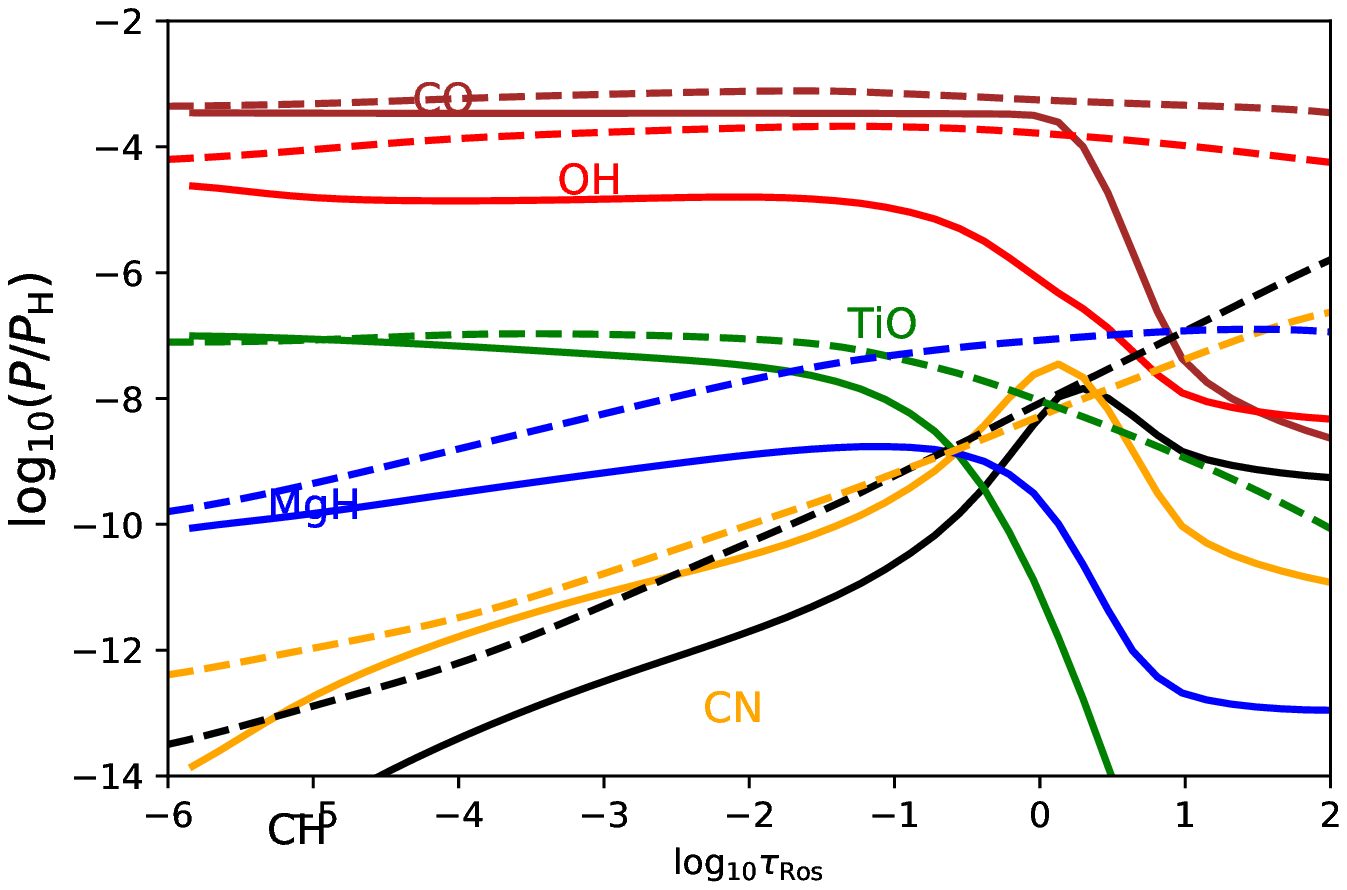}
\caption{As for Fig. \ref{fPPlogg1} for a smaller set of species, except that dashed lines are
$p_{n}$ values computed with PPRESS for a model structure converged
with PHOENIX V15.
  \label{fPHXlogg1}
}
\end{figure}

\begin{figure}
\includegraphics[width=\columnwidth]{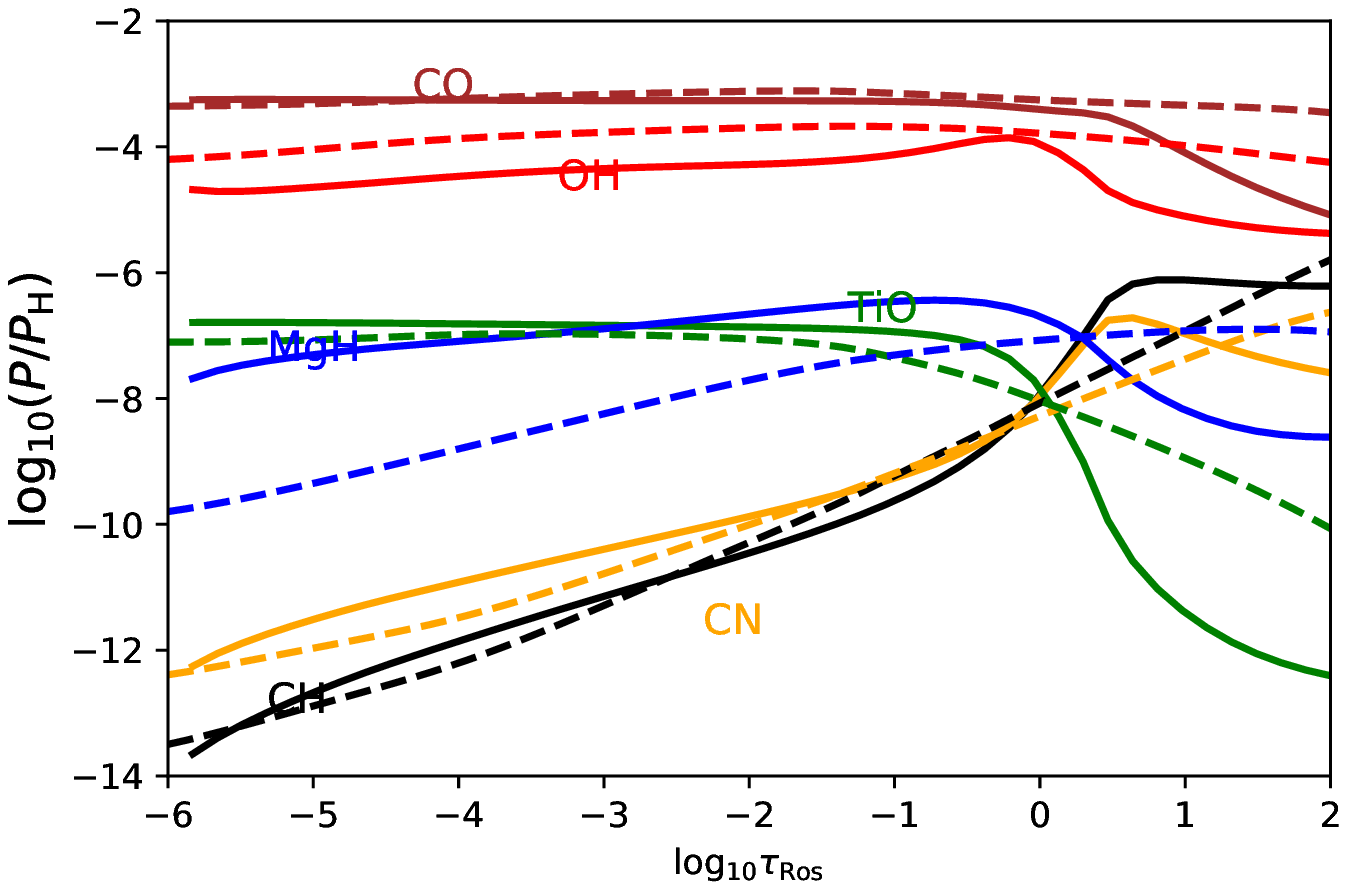}
\caption{As for Fig. \ref{fPHXlogg1} except for our model of $\log g = 5.0$.
  \label{fPHXlogg5}
}
\end{figure}

\begin{figure}
\includegraphics[width=\columnwidth]{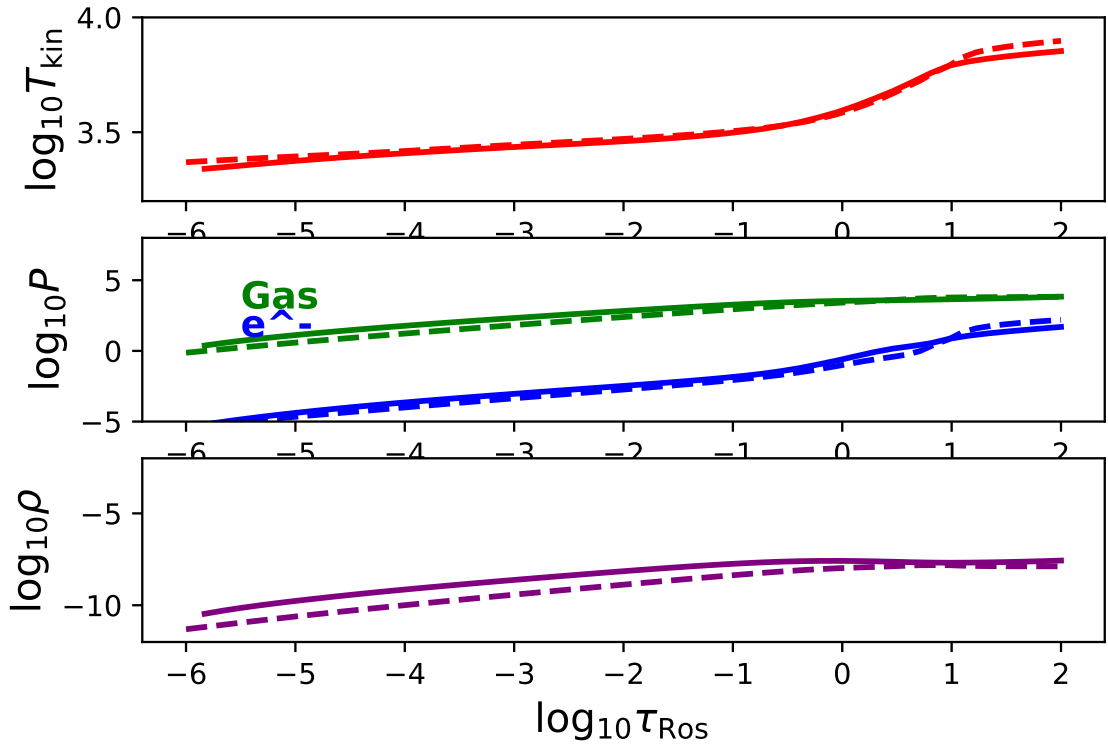}
\caption{Comparison of atmospheric structure quantities that affect the value of $p_{n}$ for the
model of $T_{\rm eff} = 3600$ K, $\log g = 1.0$ and $[{{\rm Fe}\over {\rm H}}]$ = 0.0.  Values
computed with CSPy and GAS (solid lines) and with PHOENIX V15 and PPRESS (dashed lines).
  \label{fMdlLogg1}
}
\end{figure}

\begin{figure}
\includegraphics[width=\columnwidth]{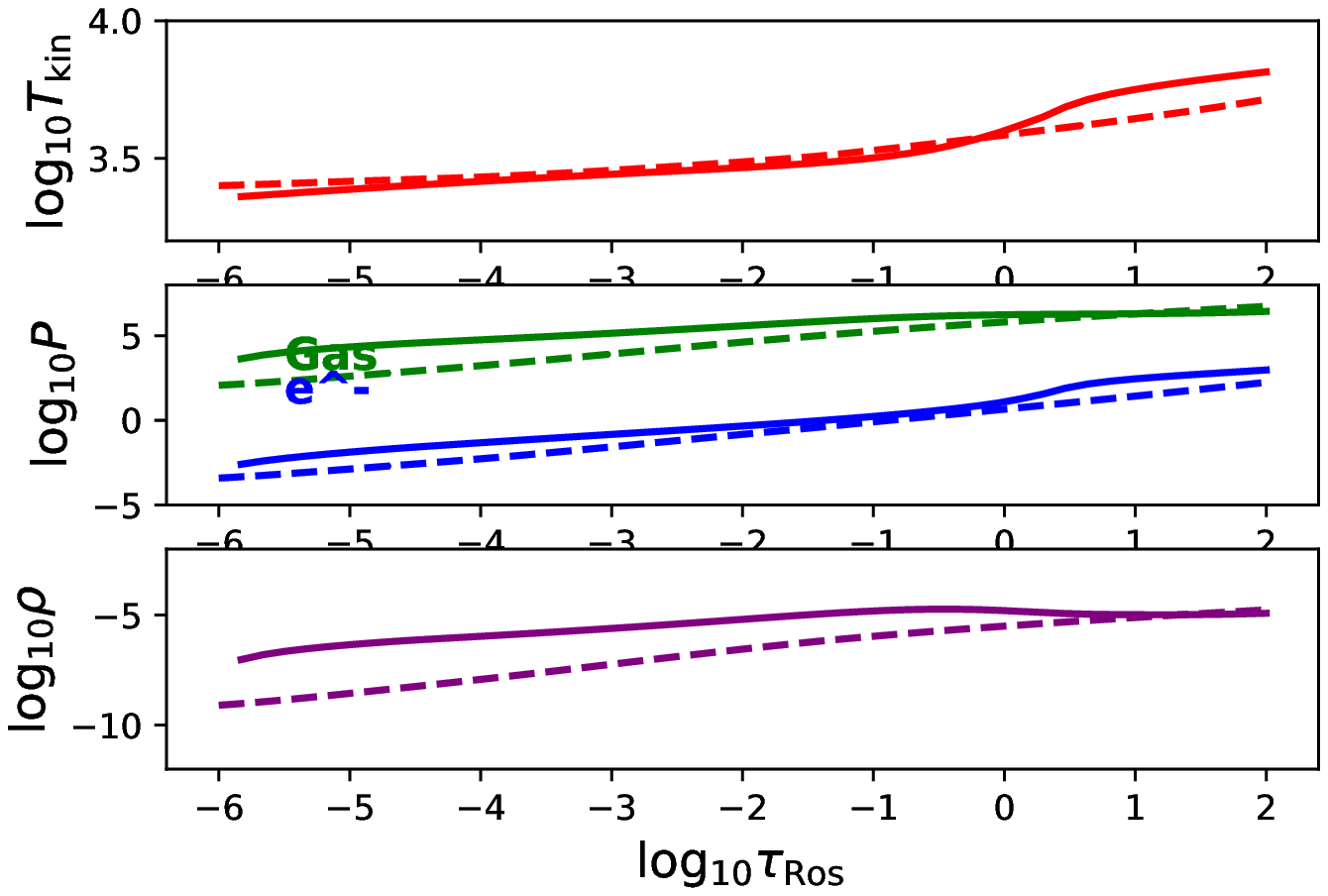}
\caption{As for Fig. \ref{fMdlLogg1}, except for a model of $\log g = 5.0$.
  \label{fMdlLogg5}
}
\end{figure}

In Figs. \ref{fPHXlogg1} and \ref{fPHXlogg5} we present a comparison of $p_{n}$ values for a smaller selection
of important molecules as computed by CSPy with GAS and by PHOENIX V15 with PPRESS, for the same stellar parameters
(3600/1.0/0.0) and (3600/5.0/0.0).   For both calculations, the abundances were those of \citet{grevs98}.  In Figs.
\ref{fMdlLogg1} and \ref{fMdlLogg5} we show the values of other state variables, $T_{\rm kin}$, $p$,
$\pe$, and $\rho$, that affect the $p_{n}$ values,
as computed with both suites.  The agreement in the $p_{n}$ values between the two packages is closest in
 the upper atmosphere ($\tau_{\rm 1200} \leq 1$) where the atmospheric structure is in radiative equilibrium,
and where $T_{\rm kin}(\tau)$
scales most closely with the value of $T_{\rm eff}$.
This is to be expected because PHOENIX computes the radiative-convective equilibrium $T_{\rm kin}(\tau)$ structure
properly throughout the entire atmosphere, whereas CSPy approximates the $T_{\rm kin}(\tau)$ structure by
re-scaling it with $T_{\rm eff}$ from one or another of three template models computed with PHOENIX V15
that sample the three populated quadrants of the HR diagram.   As a result, we expect the CSPy $T_{\rm kin}(\tau)$ structure
and, thus, the $p_{n}$ values, to be least realistic at depths of $\tau_{\rm 1200} > 1$ where the
structure is convective.


\section{Discussion and future work}
\label{discussion}

The incorporation of GAS into CSPy allows investigators, including students, to study the behavior of molecular equilibrium as a
function of $T_{\rm kin}$ and $p$ throughout stellar atmospheres of arbitrary parameters, and to do so
responsively in a Python integrated development environment (IDE).  It also allows molecular band opacity to be treated
more accurately, and adds impetus for including more molecular JOLA bands in the opacity calculation.

Because rapid responsiveness is valuable for the kinds of investigations CSPy is intended for, obtaining the $T_{\rm kin}$ structure
properly by satisfying the thermal equilibrium condition is not currently feasible, and our $T_{\rm kin}$ structure is necessarily
approximate.  Molecule formation is sensitive to the  $T_{\rm kin}$ value, so our values of $p_{n}$ for molecular species are
affected, as illustrated by the discrepancy between PHOENIX and CSPy seen in Fig. \ref{fPHXlogg5}.  A provisional measure
suggested by these results is to add an additional PHOENIX template dwarf model with a $T_{\rm eff}$ value below 4000 K for
producing scaled  $T_{\rm kin}$ structures for very late-type dwarf stars.

\paragraph{}

Our $p_{n}$ values for molecular species  are directly dependent on the quartic parameterization of $K_{n}$ of \citet{tsuji},
and these in turn affect the values for all species through the coupled chemical equilibrium.  We plan to undertake a critical
review of the molecular data in the literature with the goal of updating our treatment of $K_{n}$, and any updates will
be reported in a future paper on CSPy modelling of late-type stellar spectra.  Similarly, our treatment of the partition
function, $Q_n$, for the ionization equilibrium can be updated to reflect a more realistic $T$-dependence.

\end{document}